\documentstyle[12pt]{article}
\begin{document}
\rightline{DFPD 94/TH/30, May 1994}

\vspace{0.1cm}

{\begin{center}

{\bf DIFFUSION PROCESSES

\vspace{0.2cm}

     AND COHERENT STATES} \\

\vspace{0.5cm}

Salvatore De Martino\footnote{Electronic Mail:
demartino@vaxsa.dia.unisa.it},
Silvio De Siena\footnote{Electronic
Mail: desiena@vaxsa.dia.unisa.it}
and Giuseppe Vitiello\footnote{Electronic
Mail: vitiello@vaxsa.dia.unisa.it} \\

{\it Dipartimento di Fisica, Universit\`{a} di Salerno, and INFN, \\
     Sezione di Napoli, Gruppo collegato di Salerno, \\
     84081 Baronissi, Italia} \\

\vspace{0.2cm}

Fabrizio Illuminati\footnote{Electronic Mail:
illuminati@mvxpd5.pd.infn.it} \\

{\it Dipartimento di Fisica ``G. Galilei", Universit\`{a} di Padova, \\
     and INFN, Sezione di Padova, Via F. Marzolo 8, \\
     35131 Padova, Italia} \\

\end{center}}

\vspace{0.2cm}

{\begin{center} {\large \bf Abstract} \end{center}}

It is shown that uncertainty relations, as well
as coherent and squeezed states,
are structural properties of stochastic processes
with Fokker-Planck dynamics.
The quantum mechanical coherent
and squeezed states are explicitly constructed via Nelson
stochastic quantization. The method
is applied to derive new minimum uncertainty states
in time-dependent oscillator potentials.

\newpage

The theory of stochastic processes is the natural framework to
discuss systems with probabilistic dynamics.
As such, it is by now a most powerful tool in the study of
complex structures and behaviours in physics, chemistry, and
biology [1].
It is then very interesting to fully understand the relationship
between probabilistic and deterministic evolution, and much work
has been done in this direction [2].

In quantum mechanics, it is well known that such relationship is
expressed through uncertainty relations; in
particular, the states of minimum uncertainty, the
coherent [3] and squeezed states [4], are viewed as the
``most classical" states, closest to a deterministic time evolution.

In this letter our first observation is that uncertainty relations
analogous to the quantum mechanical ones are a structural property
of classical stochastic processes of the diffusion type.

We derive the diffusion processes
of minimum uncertainty (MUDPs), and we find that
a special class among them is associated
to Gaussian probability distributions with
time-conserved covariance
and expectation value with classical time evolution:
we denote them as strictly coherent MUDPs.
We will also consider the other Gaussian MUDPs, those with
the expectation value and the covariance in general both
time-dependent, and we will denote them as broadly coherent
MUDPs.

By exploiting Nelson's stochastic
quantization scheme [5], we will show that
MUDPs
provide the stochastic image of the standard quantum
mechanical coherent
and squeezed states, as well as of time-dependent squeezing.

Our study is motivated
by the possibility that the formalism of stochastic processes
offers to treat on the same footing, in a unified mathematical
language, the interplay between
fluctuations of different nature, for instance quantum and
thermal [6].
Our scheme holds for general
diffusion processes. Preliminary results are presented in
ref. [7].

Without loss of generality, we consider a one-dimensional random
variable $q$.
The associated diffusion process
$q(t)$ is governed by Ito's stochastic differential equation
$$
dq(t) = v_{(+)}(q(t),t)dt + {\nu}^{1/2}(q(t),t)dw(t) \, ,
\; \; \; dt > 0 \, ,
\eqno{(1)}
$$

\noindent where $v_{(+)}(q(t),t)$, is the forward drift,
$\nu (q(t),t)$ is the positive-defined
diffusion coefficient, and $dw(t)$ is a
Gaussian white noise, superimposed on the otherwise deterministic
evolution, with expectation $E(dw(t)) =0$ and covariance
$E(dw^{2}(t)) =2dt$.
The probability density $\rho (x,t)$ associated to the process
satisfies the forward and backward Fokker-Planck equations.
The forward and the backward drifts $v_{(+)}(x,t)$ and
$v_{(-)}(x,t)$ are defined as
\[
v_{(+)}(x,t) = \lim_{\Delta t \rightarrow 0^{+}}E
\left( \frac{q(t + \Delta t) - q(t)}{\Delta t}\left.
\right| q(t) = x \right) \, ,
\]
$$
\eqno{(2)}
$$
\[
v_{(-)}(x,t) = \lim_{\Delta t \rightarrow 0^{+}}E
\left( \frac{q(t) - q(t - \Delta t)}{\Delta t}\left. \right|
q(t) = x \right) \, .
\]

The operational meaning of the conditional expectations $E(\cdot \mid
\cdot)$ in eqs.(2) is the following: $v_{(+)}$ is the
mean slope of sample paths leaving point $x$ at time $t$; $v_{(-)}$
is the mean slope of sample paths entering point $x$ at time $t$.
Therefore, if $q(t)$ is a configurational process, for instance a
particle of mass $m$ performing a random motion on the real line
according to eq.(1), the forward (backward) drift is the mean
forward (backward) velocity field.

The relation between $v_{(+)}$ and $v_{(-)}$ is the following (see
also the paper by F. Guerra quoted in ref. [5]):
$$
v_{(-)}(x,t) = v_{(+)}(x,t) -
\frac{2\partial_{x}(\nu(x,t)\rho(x,t))}{\rho(x,t)} \, .
\eqno{(3)}
$$

It is  convenient to introduce the osmotic velocity $u(x,t)$ and
the current velocity $v(x,t)$
\[
u(x,t) =  \frac{v_{(+)}(x,t) - v_{(-)}(x,t)}{2} =
\frac{\partial_{x}(\nu(x,t)\rho(x,t))}{\rho(x,t)} \, ,
\]
$$
\eqno{(4)}
$$
\[
v(x,t) =  \frac{v_{(+)}(x,t) + v_{(-)}(x,t)}{2} \, .
\]

{}From the former definitions it is clear that $u(x,t)$ ``measures" the
non-differentiability of the random trajectories, thus controlling the
degree of stochasticity. In the deterministic limit $u$
vanishes and $v(x,t)$ goes to the
classical velocity $v(t)$.

Finally, we have the continuity equation
$$
\partial_{t}\rho(x,t) = -\partial_{x}(\rho(x,t)v(x,t)) \, .
\eqno{(5)}
$$

\noindent Eqs.(3)-(5) are all direct consequences of
Fokker-Planck equation.

It is straightforward to check that $E(v_{(+)}) = E(v_{(-)})
= E(v)$, $E(u)=0$, and that
$$
E(v) = \frac{d}{dt}E(q) \; \; \; \; \; \; \; \forall t \, .
\eqno{(6)}
$$

The absolute value of the expectation
of the process $q$ times the osmotic velocity $u(q,t)$ reads
$|E(qu)| = E(\nu(q,t))$. Reminding that $E(u)=0$ and by
use of Schwartz's inequality it follows that the root mean
square deviations $\Delta q = \sqrt{E(q^{2} - (E(q))^{2})}$
and $\Delta u=\sqrt{E(u^{2}-(E(q))^{2})}$ satisfy the relation
$$
\Delta q \Delta u \geq E(\nu).
\eqno{(7)}
$$

Eq.(7) is the uncertainty relation for any
stochastic process of the diffusion type defined by eq.(1).
Equality in (7) defines the MUDPs. Saturation
of Schwartz's inequality yields
$$
u(x,t) = C(t)(x - E(q)),
\eqno{(8)}
$$

\noindent with $C(t)$ an arbitrary process-independent function.

Eqs.(8) and (4) give
$$
\rho(x,t) = {\cal{N}}(t)\exp{\left[ C(t)
\int_{0}^{x}dx^{\prime}\frac{x^{\prime}
- E(q)}{\nu (x^{\prime},t)} - \ln{\nu(x,t)}\right] } \, ,
\eqno{(9)}
$$

\noindent where ${\cal{N}}(t)$ denotes the normalization function.

The density $\rho$ given by eq.(9) is associated to a large
variety of different processes. In this letter we consider
two cases: constant $\nu$ and time-dependent $\nu$.
In both cases the minimum uncertainty density is the Gaussian
$$
\rho(x,t) = \frac{1}{\sqrt{2\pi(\Delta q)^{2}}}
\exp{\left[-\frac{(x - E(q))^{2}}{2(\Delta q)^{2}}\right]} \, ,
\eqno{(10)}
$$

\noindent with $(\Delta q)^{2}=-\nu(t)/C(t)$.

The continuity equation, eq.(5), forces the current velocity
$v(x,t)$ to be of the form
$$
v(x,t) = \frac{d}{dt}E(q) + \left( x-E(q) \right)
\frac{d}{dt}\ln{\Delta q} \, .
\eqno{(11)}
$$

The stochastic differential equation obeyed by the MUDPs is now
completely determined by eqs.(10)-(11) and it reads
$$
dq(t) = \left[ A(t) + B(t)q(t) \right] dt + \nu^{1/2}(t) dw(t) \, .
\eqno{(12)}
$$

\noindent It is interesting to observe that eq.(12) has a drift
part which is linear in the process. The above equation in fact
defines the so-called {\it linear processes in narrow sense}
[1]; when $A(t)=0$ they are the time-dependent Ornstein-Uhlenbeck
processes.

We can now divide the MUDPS in two general
classes, the first one with
$$
\left\{ \begin{array}{ll}
\Delta q  & =  \mbox{const.} \; \; \; \; \; \; \forall t \, , \\
E(q) & = a(t) \, , \end{array}
\right.
\eqno{(13a)}
$$

\noindent and the second one with
$$
\left\{ \begin{array}{ll}
\Delta q & =  F(t) \, , \\
E(q) & = b(t) \, , \end{array}
\right.
\eqno{(13b)}
$$

\noindent where $a(t)$, $b(t)$ and $F(t)$ are arbitrary functions
of time.

In the case (13a), $\Delta q$
does not spread and  $E(q)$ follows a classical
trajectory:
$$
v(x,t)=\frac{d}{dt}E(q) = v(t) \, .
\eqno{(14)}
$$

As a consequence, MUDPs of the
form (10) obeying eqs.(13a)-(14) behave exactly as the
quantum mechanical coherent
states: we will denote them as {\it strictly} coherent MUDPs,
as opposed to the ones obeying eq.(13b) which we call
{\it broadly} coherent MUDPs.

It is possible to discriminate on physical grounds the strictly
coherent MUDPs from the other MUDPS by observing
that eqs.(13a)-(14) are immediate consequences of the
Ehrenfest condition
$$
v(E(q),t)=\frac{d}{dt}E(q) \, ,
\eqno{(15)}
$$

\noindent so that the strictly coherent MUDPs can be viewed
as the most deterministic semi-classical processes.

We shall now consider a very important class of conservative
diffusion processes (Nelson diffusions) which has been introduced
by Nelson in the stochastic formulation of quantum mechanics [5].

Nelson stochastic quantization associates to each
single-particle quantum state
$\Psi = \exp{ [ R +
\frac{i}{\hbar}S ] }$, the diffusion process $q(t)$ with

$$
\nu=\frac{\hbar}{2m} \, , \; \;
\rho(x,t)=|\Psi(x,t)|^{2}\, ,  \; \;
v(x,t) = \frac{1}{m}\frac{\partial S(x,t)}{\partial x} \, ,
\eqno{(16)}
$$

\noindent where $m$ is the mass of the particle, and
$R$ is related to $\rho$ by the obvious relation
$\rho=\exp [2R]$.

The Schr\"{o}dinger equation with potential
$V(x,t)$ leads to the
Hamilton-Jacobi-Madelung equation
$$
{\partial}_{t}S(x,t) + \frac{({\partial}_{x} S(x,t))^{2}}{2m}
- \frac{\hbar^{2}}{2m}\frac{\partial_{x}^{2}\rho^{1/2}(x,t)}
{\rho^{1/2}(x,t)} = -V(x,t) \, .
\eqno{(17)}
$$

It is well known [8] that for Nelson diffusions
the correspondences between the expectations of
stochastic and operatorial observables are
\[
\langle \hat{q} \rangle = E(q) \, ,
\]

\[
\langle \hat{p} \rangle = mE(v) \, ,
\]

$$
\Delta \hat{q} = \Delta q \, ,
\eqno{(18)}
$$

\[
(\Delta \hat{p})^{2} = m^{2}[(\Delta u)^{2} + (\Delta v)^{2}] \, ,
\]

\[
(\Delta \hat{q})^{2} (\Delta \hat{p})^{2} \geq
m^{2}(\Delta q)^{2} (\Delta u)^{2} \geq \frac{{\hbar}^{2}}{4} \, ,
\]

\noindent where the hat denotes the operatorial observables,
$\langle \cdot \rangle$ denotes the expectation value in a
given state $\Psi$ and $\Delta (\cdot)$ denotes
the root mean square deviation.

Minimum uncertainty Nelson diffusions (MUNDs) are  MUDPs,
and we correspondingly extend to them the
denominations of
strictly and broadly coherent MUNDs.
Relation (17) can be regarded as an equation for
the potential $V(x,t)$.
In the case of strictly coherent MUNDs (13a), we obtain
$$
V(x,t) = \frac{m}{2}\omega^{2}x^{2} + f(t)x + V_{0}(t) \, ,
\eqno{(19a)}
$$
$$
\frac{d^{2}}{dt^{2}}E(q) =
- \omega^{2}E(q) + \frac{f(t)}{m}  \, ,
\eqno{(19b)}
$$

\noindent with $f(t)$ and $V_{0}(t)$ arbitrary time-dependent
functions and the constant frequency
$$
\omega^2 = \frac{\hbar^2}{4m^{2}(\Delta q)^4} \, .
\eqno{(20)}
$$

For broadly coherent MUNDs (12b) we again obtain eqs.(19a)-(19b)
but now with a frequency $\omega(t)$ depending on time through the
spreading $\Delta q$:
$$
\omega^{2}(t) = \frac{\hbar^{2}}{4m^{2}(\Delta q)^{4}} -
\frac{1}{\Delta q}\frac{d^{2}}{dt^{2}}\Delta q \, .
\eqno{(21)}
$$

We now address the case of time-dependent $\nu$. From
the first of eqs. (16) this means
letting either $m$ or $\hbar$, or both, be functions of time.

We focus our attention on the case of time-dependent mass $m(t)$
and constant $\hbar$, leaving apart
other more speculative situations.

The Nelson scheme (16)-(17)
still holds with $m(t)$ replacing $m$.
Considering the general case of broadly coherent MUNDs,
and solving eq.(17) for $V(x,t)$ we obtain
\[
V(x,t) = \frac{1}{2}m(t)\omega^{2}(t)x^{2} +
f(t)x + V_{0}(t) \, ,
\]
$$
\omega^{2}(t) = \frac{\hbar^{2}}{4m^{2}(t)(\Delta q)^{4}} -
\frac{1}{\Delta q}\frac{d^{2}}{dt^{2}}\Delta q
-\frac{\dot{m}(t)}{m(t)\Delta q}\frac{d}{dt}\Delta q \, ,
\eqno{(22)}
$$

\[
\frac{d^{2}}{dt^{2}}E(q) =
- \frac{\dot{m}(t)}{m(t)}\frac{d}{dt}E(q)
- \omega^{2}(t)E(q) + \frac{f(t)}{m(t)} \, ,
\]

\noindent where $f(t)$, $V_{0}(t)$ are arbitrary functions of time
and $\dot{m}$ denotes the time derivative of $m$. The subcase of
strictly coherent MUNDs for systems with a time-dependent mass
is obviously recovered by putting $\Delta q = const.$ in eqs.(22).

To illustrate the advantages of the formalism presented in this
paper we observe that it is computationally convenient in many
cases of practical interest.
For instance, when the arbitrary functions $f(t)$
and $V_{0}(t)$ vanish, eqs.(19a)-(19b) are those of
the classical harmonic
oscillator, and the associated quantum states are the standard
Glauber coherent states; when $f(t)=const.$ we have the
Klauder-Sudarshan displaced oscillator coherent states; finally,
when
$f(t)$ is time-dependent, we obtain the Klauder-Sudarshan driven
oscillator coherent states [9]. We thus see that this formalism
at once provides the full set of coherent states so widely exploited
in physical applications.

Moreover, we observe that eqs.(19a)-(19b) supplemented by eq.(21)
describe the dynamics of the parametric
oscillator with the associated feature of time-dependent
squeezing. Finally,
eqs.(22), supplemented with $m(t)=m_{0}e^{\Gamma(t)}$, define the
dynamics of the damped parametric oscillator, a result which sheds
new light on the study of dissipative quantum mechanical
systems ([10], [11])
also in view of the relation among squeezing and dissipation [12].

Besides the computational simplicity, we
would like to stress that the stochastic formalism also provides
new insights beyond the conventional operatorial framework:
it is in fact most remarkable that coherent and squeezed states
of different types, time-dependent oscillators and dissipative
systems may all be described
in terms of, and associated to, diffusion processes
via Nelson stochastic quantization.

We note that eqs.(16) and (17) not only yield the
potential of the quantum state associated to the MUND, but also
allow
to compute explicitly the wave function. In the following we
exhibit two cases; the first is the case of the familiar
Glauber state, the second one is the case of the coherent
state associated to a dissipative dynamics of the Caldirola-Kanai
type.

The wave function for the Glauber state which we indeed
obtain from
eqs.(10),(14) and (17), together with the maps (16) and (18),
is
$$
\Psi_{G}(x,t)={\frac{1}{(2 \pi (\Delta {\hat q})^2)^{\frac{1}{4}}}}
\exp \left[ -{\frac{( x -\langle \hat{q} \rangle )^{2}}
{4(\Delta {\hat q})^{2}}}
+{\frac{i}{\hbar}} x \langle \hat{p} \rangle \right] \, \, ,
\eqno{(23)}
$$

\noindent where $\langle \hat{q} \rangle (t)$ and
$\langle \hat{p} \rangle (t) = mv(t) =
m(d \langle \hat{q} \rangle /dt)$ are the solutions of
the classical equations of motion of the harmonic
oscillator (see eqs.(19a)-(19b) with the choice $f(t)=V_{0}(t)=0$);
the wave funtion (23) describes, as it is well known, both
the coherent and the squeezed states, since it
is form invariant under the
scale transformation ${\hat q} \rightarrow e^{s} {\hat q}$
and ${\hat p} \rightarrow e^{-s} {\hat p}$.

For the more intriguing case
of the damped parametric oscillator, we obtain, through the
same procedure, the wave function
\[
\Psi_{D}(x,t)= {\frac{1}{(2 \pi
(\Delta {\hat q})^{2})^{\frac{1}{4}}}}
\exp \left\{ -{\frac{(x-\langle \hat{q} \rangle )^{2}}
{4 (\Delta {\hat q})^{2}}}
+{\frac{i}{\hbar}} \left[ x \langle \hat{p} \rangle + \frac{ \langle \hat{q}
\hat{p} \rangle - \langle \hat{q} \rangle \langle \hat{p} \rangle }
{2(\Delta \hat{q})^{2}}(x- \langle \hat{q} \rangle )^{2}
\right] \right\} \, ,
\]
$$
\eqno{(24)}
$$

\noindent where now $\langle \hat{q} \rangle (t)$, and
$\langle \hat{p}\rangle (t)=m(t)v(t)=m(t)(d\langle
\hat{q}\rangle /dt)$ are the solutions of the classical
equation of motion
of the damped parametric oscillator (see eqs.(22) where
we have put for simplicity $f(t)=V_{0}(t)=0$), and
we have exploited the property, easy to verify, that
$$
m(\Delta q)^{2}\frac{d}{dt}\ln{\Delta q} = m\left[
E(qv) - E(q)E(v) \right] = \langle \hat{q} \hat{p} \rangle
- \langle \hat{q} \rangle \langle \hat{p} \rangle \, .
\eqno{(25)}
$$

Equation (25) yields the correspondence between the stochastic
and the operatorial correlations among the quantum observables;
by exploiting it, we have that
$$
m\left[E(qv) - E(q)E(v) \right] =
{\frac{\langle \{ {\hat Q},{\hat P} \} \rangle }{2}} \, ,
\eqno{(26a)}
$$

\noindent where
$$
\hat{Q} = \hat{q} -\langle \hat{q} \rangle \, \; , \; \;
\; \; \; \;
\hat{P} = \hat{p} -\langle \hat{p} \rangle \, .
\eqno{(26b)}
$$

By exploiting the above relations (26a)-(26b)
we can immediately verify that the states
corresponding to eq.(20) are
Heisenberg minimum uncertainty (m.u.) states,
and those corresponding to
eq.(21) are Schr\"odinger ones. Indeed, one can prove
that the strictly coherent MUNDs are in one to one correspondence
with the Heisenberg m.u. states, while broadly coherent MUNDs are
all and only Schr\"odinger m.u. states.

The m.u. states already known
for the damped parametric oscillator are Schr\"odinger ones
and there was in the literature a widespread belief that
this physical system cannot have Heisenberg m.u. states [10];
by eqs.(24)-(26) we see instead that it
can also exhibit Heisenberg m.u. states, strictly coherent
and harmonic oscillator-like. Thus, for the dissipative
oscillator the stochastic approach
allows to determine not only all the known
Schr\"odinger m.u. states, but also a whole new set of
Heisenberg m.u. states.

In conclusion it seems to us interesting and stimulating
the possibility to relate
coherent states with diffusion processes and probabilistic
methods, which provide the proper definition of
functional integration techniques, and
the appropriate framework for
the study of dissipative systems.

\newpage

{\begin{verse} {\Large {\bf References}} \\
\vspace{1cm}
[1] See e.g.:
C. W. Gardiner, {\it Handbook of Stochastic Methods} (Springer,
Berlin, 1985). \\
\vspace{0.6cm}
[2] {\it Stochastic Processes in Classical and Quantum
Systems}, edited by S. Albeverio, G. Casati, and D. Merlini
(Lecture Notes in Physics no.262, Springer, Berlin, 1986). \\
\vspace{0.6cm}
[3] See e.g.: J. R. Klauder and B. S. Skagerstam, {\it Coherent
States}
(World Scientific, Singapore, 1985). \\
\vspace{0.6cm}
[4] D. Stoler, Phys. Rev. {\bf D1}, 3217 (1970); H. P. Yuen, Phys.
Rev. {\bf A13}, 2226 (1976). \\
\vspace{0.6cm}
[5] E. Nelson {\it Dynamical Theories of Brownian Motion}
(Princeton University Press, Princeton, 1967);
{\it Quantum Fluctuations} (Princeton University
Press, Princeton, 1985); F. Guerra, Phys. Rep. {\bf 77}, 263 (1981);
G. Parisi, {\it Statistical Field Theory}
(Addison-Wesley, New York, 1988). \\
\vspace{0.6cm}
[6] P. Ruggiero and M. Zannetti, Phys. Rev. Lett. {\bf 48},
963 (1982). \\
\vspace{0.6cm}
[7] S. De Martino, S. De Siena, F. Illuminati, and G. Vitiello,
in {\it Proceedings of the Third International
Workshop on Squeezed States and Uncertainty Relations},
edited by D. Han {\it et al.} (NASA Conference Publication
3270, NASA, Greenbelt MD, 1994), 331-336. \\
\vspace{0.6cm}
[8] L. De La Pe\~{n}a and A. M. Cetto,
Phys. Lett. {\bf A39}, 65 (1972); D. de Falco, S. De Martino,
and S. De Siena, Phys. Rev. Lett. {\bf 49}, 181 (1982);
S. De Martino and S. De Siena, Nuovo Cimento {\bf B79},
175 (1984). \\
\vspace{0.6cm}
[9] J. R. Klauder and E. C. G. Sudarshan, {\it Fundamentals
of Quantum Optics} (Benjamin, New York, 1970). \\
\vspace{0.6cm}
[10] O. V. Man'ko, in {\it Proceedings of the Workshop on
Harmonic Oscillators}, edited by D. Han {\it et al.}
(NASA Conference Publication 3197, NASA, Greenbelt MD, 1992). \\
\vspace{0.6cm}
[11] E. Celeghini, M. Rasetti and G. Vitiello, Ann. of Phys.
{\bf 215}, 156 (1992). \\
\vspace{0.6cm}
[12] E. Celeghini, M. Rasetti, M. Tarlini and G. Vitiello,
Mod. Phys. Lett. {\bf B3}, 1213 (1989). \\
\vspace{0.6cm}
\end{verse}}

\end{document}